\documentclass[twocolumn,amsmath,longbibliography,amssymb,superscriptaddress]{revtex4-2}

\usepackage[pdftex]{graphics}
\usepackage{graphicx}
\graphicspath{{figures/}}
\usepackage{hyperref}
\usepackage{xcolor}
\usepackage{physics}
\usepackage{subfig}
\usepackage{bm}
\usepackage{caption}
\usepackage{amsmath} 
\usepackage{mathtools}

\begin{document}
		
\title{From Hermitian critical to non-Hermitian point-gapped phases}
\author{Carlos Ortega-Taberner}
\affiliation{Department of Physics, Stockholm University, AlbaNova University Center, SE-106 91 Stockholm, Sweden}
\affiliation{Nordita, KTH Royal Institute of Technology and Stockholm University, SE-106 91 Stockholm, Sweden}

\author{Maria Hermanns}
\affiliation{Department of Physics, Stockholm University, AlbaNova University Center, SE-106 91 Stockholm, Sweden}
\affiliation{Nordita, KTH Royal Institute of Technology and Stockholm University, SE-106 91 Stockholm, Sweden}

\date{\today}
\begin{abstract}
Recent years have seen a growing interest in topological phases beyond the standard paradigm of gapped, isolated systems. One recent direction is to explore topological features in non-hermitian systems that are commonly used as effective descriptions of open systems. Another direction explores the fate of topology at critical points, where the bulk gap collapses. One interesting observation is that both systems, though very different, share certain topological features. 
For instance, both systems can host half-integer quantized winding numbers and have very similar entanglement spectra. Here, we make this similarity explicit by showing the equivalence of topological invariants in critical systems with non-hermitian point-gap phases, in the presence of sublattice symmetry. This correspondence may carry over to other features beyond topological invariants, and may even be helpful to deepen our understanding of non-hermitian systems using our knowledge of critical systems, and vice versa. 
\end{abstract}

\maketitle

\section{Introduction}

In recent years non-Hermitian systems have attracted a lot of attention in the condensed-matter community due to the unique phenomena that they can exhibit.
In the field of symmetry protected topology, non-Hermitian systems harbor a particularly rich variety of phases, as the non-Hermiticity enhances the ten Hermitian topological classes \cite{Altland1997,schnyder_classification_2008,Ludwig2015} into 38 \cite{Bernard2002, Gong2018, PhysRevX.9.041015,PhysRevB.99.235112}. 
This enhancement originates in having effectively more symmetries available, as conjugation and transposition are not equivalent any longer. 
Furthermore, non-Hermiticity may lead to new physics without a Hermitian equivalent: 
So-called point-gap phases and exceptional points \cite{bergholtz_exceptional_2021, Ashida_non_2020} lead to new physics with no Hermitian equivalent, like the skin effect \cite{Gong2018,Okuma2020,bergholtz_exceptional_2021,okuma_quantum_2021,edvardssonSensitivityNonHermitianSystems2022}.

For non-Hermitian systems, we need to distinguish between line-gapped and point-gapped systems. 
For the former, it is possible to draw a line in the complex plane which separates the different bands. 
On the other hand, a phase is said to be point-gapped at $E_0$ if it has a non-zero `energy vorticity' for this point, i.e. the complex energy bands wind around $E_0$. 
While the topology of line-gapped systems can be understood from their connection to Hermitian and anti-Hermitian Hamiltonians \cite{PhysRevX.9.041015}, point-gapped systems have features that are thought to be intrinsically  non-Hermitian. 
In this paper we will focus on these systems and show how some of their features are inherited from related Hermitian critical models.

The topic of topological phenomena in critical systems has recently been revisited \cite{verresen2018,verresen2020,VerresenPRX,BEH,Balabanov_2022}. Topology is a global (non-local) property of the system, and it was previously thought that the divergent correlation length of critical systems always rendered the system topologically trivial. However, it has been shown recently that this is not the case \cite{verresen2018}. Critical systems can harbor topologically protected zero-energy edge states that are protected by a topological invariant and do not hybridize with the bulk. 
The topological phases in these systems have been characterized by invariants that are quantized to half-integers \cite{verresen2020}. 

The latter is also true for some non-Hermitian systems. For example, systems in the topological class AI (with a sublattice symmetry S that commutes with time-reversal), are characterized by two winding numbers which can take half-integer values when the system has a point-gap \cite{Yin2018,Gong2018,Jian_winding_2018,masuda_relationship_2021}. This similarity in the topology of Hermitian critical  and non-Hermitian systems has been pointed out in the literature before \cite{Sarkar2018}, but to our knowledge it has not yet been explored in depth. 
In this paper we aim to relate the bulk topological features of critical Hermitian systems and point-gapped non-Hermitian systems. 
We will introduce two different methods of generalizing a Hermitian critical Hamiltonian to a non-Hermitian one, showing that the resulting model always is in a point-gap phase.  
We also analyze how bulk topological features evolve with the non-Hermiticity, in particular the topological invariants and the so-called entanglement occupancy spectrum (EOS) \cite{Peschel1999,Peschel2003,Peschel2008}. 
While the two generalization considered here are somewhat limited, the second includes two of the most used non-Hermitian models, the Hatano-Nelson model and the SSH chain with unbalanced hopping.

In Hermitian systems the EOS, computed for periodic boundary conditions, provides the same topological information as the surface energy spectrum \cite{Fidkowski2010entanglement,yao2010entanglement}, and it has been connected to topological invariants such as the polarization \cite{ortega-taberner_relation_2021} and the winding number \cite{Monkman2022}. 
For non-Hermitian point-gapped systems the bulk-boundary correspondence is broken due to the skin effect.
Thus one cannot use the surface spectrum to study bulk topology (and vice-versa). 
In this case, the EOS becomes more interesting, as it can be computed for periodic boundary conditions and can be related to bulk invariants \cite{herviou_defining_2019, herviou_entanglement_2019}.
Connecting these two, on first glance unrelated types of phases, would allow one to compute objects for non-Hermitian models using Hermitian physics, as well as use the more developed topological classification for non-Hermitian systems to address questions regarding critical Hermitian systems. 

The paper is organized as follows: In section \ref{sec:Intro} we cover some necessary background material. 
In section \ref{sec:generalization_1}, we consider a very simple non-Hermitian generalization to elucidate the main ideas. 
A more complex generalization is shown in section \ref{section:complex_momenta}, where the non-Hermiticity is introduced by making the momenta complex. 
The entanglement occupancy spectrum is discussed in  section \ref{sec:EOS}. 
Some explicit examples are covered in Appendix \ref{sec:examples}.

\section{Background}\label{sec:Intro}

Throughout the paper, we will restrict the discussion to a general 2-band model given by the Hamiltonian 
\begin{eqnarray}\label{eq:Hamiltonian}
    H(k)&=& \boldsymbol{h}(k)\cdot \sigma \nonumber\\
    & =& h_x(k)\sigma_x + h_y(k) \sigma_y \nonumber\\
    &=& \left( \begin{array}{cc}
    0 & f_1 (k) \\ f_2 (k) & 0
    \end{array}
    \right),
\end{eqnarray}
where we set $h_z$ to zero because of sublattice symmetry.  Both ways of describing the Hamiltonian will prove useful in the following. 
Note that for non-Hermitian models, $f_2(k)\neq f_1^*(k)$, while in the Hermitian case we will suppress the index, and use $f(k) \equiv f_1(k) = f^*_2 (k)$.
The right eigenstates of such two-band Hamiltonians are in general given by 
\begin{align}\label{eq:eigenstates}
    v^R_{\pm}(k) &= \frac{1}{\sqrt{2}}(\pm \sqrt{f_1(k)/f_2(k)},1)
\end{align}
with energies $\pm \sqrt{f_1(k)f_2(k)}$. 
The corresponding left ones are 
\begin{align}\label{eq:eigenstates_left}
    v^L_{\pm}(k) &= \frac{1}{\sqrt{2}}(\pm \sqrt{f_2(k)/f_1(k)},1)
\end{align}

For Hermitian two-band models, the winding number \cite{Ludwig2015} can be written as 
\begin{align}\label{eq:winding_number_arg}
    \nu &= \frac{1}{2\pi} \int_0^{2\pi} dk \, \partial_k \arg[h_x(k)-i h_y(k)], 
\end{align}
or alternatively as 
\begin{align}\label{eq:winding_number_nH}
   \nu & =\frac{1}{2\pi}\int_0^{2\pi}dk\, \frac{(\partial_k h_x(k) ) h_y(k)-h_x(k) \partial_k h_y(k)}{ h_x(k)^2+h_y(k)^2}. 
\end{align}
For non-Hermitian models both expressions are no longer equal. The second expression, Eq.~\eqref{eq:winding_number_nH}, is usually employed to compute the winding number \cite{Yin2018,Gong2018,PhysRevX.9.041015}.  
While the Hamiltonian \eqref{eq:Hamiltonian} is, of course, restrictive, it nevertheless covers two of the most studied non-Hermitian models, the Hatano-Nelson model \cite{Hatano1996,Hatano1997,Hatano1998} and the non-Hermitian extension to the SSH model \cite{Yao2018}.

\subsection{Topological invariants for critical systems}
Topological invariants are usually considered meaningful only in gapped systems.
This raises the question if and how topological features survive the presence of gapless modes, e.g. by coupling to a gapless environment \cite{bondersonQuasitopologicalPhasesMatter2013} or by driving to a phase transition between two topological phases \cite{verresen2018, verresen2020}. 
For the latter, the usual definitions of topological invariants become ill-defined. 
A general `recipe' on obtaining well-defined invariants for critical systems was first presented in \cite{verresen2020}, by removing an infinitesimal region around each of the gapless points. 

For the winding number, this implies that instead of \eqref{eq:winding_number_arg}, we use the following, regularized version:
\begin{align}\label{eq:winding_number_epsilon}
    \nu_\epsilon &= \frac{1}{2\pi} \int_{()} dk \, \partial_k \arg[h_x(k)-i h_y(k)]
\end{align}
where $()$ denotes that  we removed a region $[k_c-\epsilon,k_c+\epsilon]$ from the integration around each gapless point $k_c$. 
In the limit $\epsilon \rightarrow 0$, $\nu_\epsilon$ becomes half-integer quantized, and is given generically by the mean of the winding numbers on either side of the critical point, for details see Ref.~\cite{verresen2020}.   
In particular, at the critical point between gapped phases with winding number 0 and 1 respectively, the winding number $\nu_\epsilon \rightarrow  1/2$ for $\epsilon\rightarrow 0$. 

There is an alternative formulation of the winding number by interpreting $f (=f_1=f_2^*)$ in Eq.~\eqref{eq:Hamiltonian} as a complex function, $f(z=e^{ik})$, and relating the winding number to the difference in numbers of zeroes and poles within the unit circle, using Cauchys Theorem \cite{verresen2018,BEH},
\begin{align}
	\nu &= \frac{1}{2\pi i} \oint dz \frac{f'(z)}{f(z)} = Z-P,
\end{align}
where $Z/P$ denotes the number of zeroes/poles within the unit circle. 
 This identity is not valid for gapless systems because there is a zero in the contour. 
One alternative for gapless systems is to again exclude an infinitesimal region around each point $z$,
\begin{align}
\nu_\epsilon &= \frac{1}{2\pi i} \oint_{()} dz \, \frac{f'(z)}{f(z)}.
\end{align}
Each zero of order $n$ on the unit circle contributes $n/2$ to the winding number, in agreement with the discussion in the section above.
One could, however, have chosen to define the invariant differently, namely as $\tilde{\nu}=Z-P$, where $Z$/$P$ are the zeroes and poles strictly within the unit circle. This instead gives an integer (not half-integer) quantized invariant, which in addition is related to the number of topologically protected edge modes \cite{verresen2018,BEH}.
  
One disadvantage of Verresen's approach in Ref.~\cite{verresen2020} is that one needs to compute different invariants, depending on whether the system is gapless or gapped --- the invariant for gapped systems is ill-defined for gapless ones. 
A different way of regularizing the winding number for gapless systems was discussed in  Ref.~\cite{Balabanov_2022}, namely by considering the system at finite temperatures.
Computing the $T\rightarrow 0$ limit in a gapped system yields \eqref{eq:winding_number_arg}, while doing the same for a critical system yields the $\epsilon \rightarrow 0$ limit of Eq.~\eqref{eq:winding_number_epsilon}, as the Boltzmann weights suppress the contribution in the vicinity of gapless points. It, thus, has the advantage that gapped and gapless systems can be treated on the same footing. 
The generalization to non-Hermitian systems detailed in sections \ref{sec:generalization_1} and \ref{section:complex_momenta} can, in some sense, be regarded as yet another regularization of the winding number that treats gapped and gappless systems on the same footing.  

Note that the winding number $\lim_{\epsilon \rightarrow 0}\nu_{\epsilon}$ does not uniquely characterize the critical system, nor is it connected to the number of topologically protected edge modes. 
Two different critical points --- e.g. the phase transitions  $\nu=0\rightarrow 3$ and $\nu=1\rightarrow 2$ --- have the same value $\nu=3/2$ of the winding number. 
The first critical point has no protected topological egde modes, while the latter has one. 
In general, the number of topological edge modes is given by \cite{verresen2018}
\begin{align}\label{eq:number_top_edge_modes}
    N_{\mbox{\tiny top}} =\left\{ \begin{array}{cc}
        \tilde\nu & \mbox{ if $\tilde\nu>0$. }  \\
        |\tilde\nu+N| & \mbox{ if $\tilde\nu<-N$}\\
        0 & \mbox{ otherwise}
    \end{array}\right.,
\end{align}
where $\tilde\nu=Z-P=\nu -  N/2$, $Z/P$ are the number of zeroes/poles strictly within the unit circle, and $N$ is the number of zeroes (times their multiplicity) on the unit circle. 
A complete characterization of the critical system thus requires two  numbers, e.g.  $\tilde \nu$ and $N$. 
This is very similar to the non-Hermitian, point-gap phases discussed below.

\subsection{Non-Hermitian systems with sublattice symmetry}
In this paper, we focus on systems in symmetry class AI, with a sublattice symmetry S that commutes with time-reversal.\footnote{In the notation of Table VII in reference \cite{PhysRevX.9.041015} we consider systems in AI with S$_+$.}
Such systems are specified by two winding numbers. 
A commonly used characterization uses winding numbers $\nu$ and $\nu'$, where $\nu$ is given by Eq. \eqref{eq:winding_number_nH} and $\nu'$ denotes the winding of the complex energy bands around the origin \cite{Gong2018,Okuma2020} (sometimes referred to as `energy vorticity'):
\begin{align}
    \nu'&=\frac{1}{2\pi}\oint dk \, \partial_k \arg\left( \sqrt{h_x(k)^2+h_y(k)^2}\right). 
\end{align}
An alternative formulation was presented in Ref.~\cite{Yin2018}.
Since $f_1\neq f_2^*$ \eqref{eq:Hamiltonian} in non-Hermitian systems, we can define two independent winding numbers by 
\begin{align}\label{eq:nu1/2}
	\nu_1 &= \frac{1}{2\pi i} \int dk \, \partial_k  \ln[f_1(k)] \nonumber\\
	\nu_2 &=- \frac{1}{2\pi i} \int dk \, \partial_k \ln[f_2(k)], 
\end{align}
which fulfill
\begin{align}\label{eq:winding2}
&\nu = \frac{1}{2}(\nu_1+\nu_2) \nonumber\\
&\nu' = \frac{1}{2}(\nu_1-\nu_2). 
\end{align}
Note that the `$-$' sign in the second line of Eq.~\eqref{eq:nu1/2} ensures that $\nu_1=\nu_2$ in the Hermitian limit. \footnote{We should note that our conventions differ from those of \cite{Yin2018}. Using a tilde to indicate the conventions of the latter, we use $\nu=-\tilde\nu$, $\nu'=\tilde \nu'$, $\nu_1=-\tilde \nu_2$ and $\nu_2=-\tilde \nu_1$. }
An appealing feature of the winding numbers $\nu_{1/2}$ is that they are directly related to the presence of topologically protected boundary modes present at the right/left edge of a semi-infinite open system \cite{Yin2018}. 
We will later see that $\nu_{1/2}$ also have a natural interpretation in the connected Hermitian, critical system: they can often be identified with the winding numbers of the neighboring gapped phases, which will be discussed in Section \ref{section:complex_momenta}.

\subsection{Entanglement occupancy spectrum}\label{sec:EOS_intro}
The entanglement spectrum was originally proposed for strongly interacting fractional quantum Hall liquids \cite{Li2008entanglement} as a usefull tool to obtain information about the topology of the state. 
It is obtained by first partitioning the system into two parts A and B (usually in real space) and tracing out the degrees of freedom in B to obtain the reduced density matrix $\rho_A = \mbox{Tr}_B(|\psi\rangle\langle \psi|)$, where $|\psi\rangle$ denotes the ground state of the system. 
We can now define an entanglement Hamiltonian $H_A$ by 
\begin{align}
    \rho_A&= e^{-H_A}, 
\end{align}
whose spectrum is called entanglement spectrum and contains valuable information about the topological properties of the system. 

The entanglement spectrum is also of interest in non-interacting systems, as it is directly connected to the topological edge modes in gapped topological insulators \cite{Fidkowski2010entanglement}.
 Peschel showed that for non-interacting systems one does not need to compute the full many-body entanglement spectrum (a difficult task even for non-interacting systems)~\cite{Peschel2003}.
Instead, one can obtain the same information from the spectrum of the reduced correlation matrix
\begin{align}\label{eq:CA_gen}
    C^A_{j,j',\alpha,\beta} &= \langle gs | c_{j,\alpha}^\dagger c_{j',\beta}|gs\rangle ,
\end{align}
where $\alpha$ and $\beta$ denote internal degrees of freedom, and $j,j'$ denote sites in region A. 
In the following, we will call the spectrum of $C^A$ the entanglement occupancy spectrum (EOS). 
Topological states correspond to $\xi=1/2$ modes of the EOS, since these give rise to degeneracies in the many-body entanglement spectrum.
The number of $\xi=1/2$ modes is identical to the number of topologically protected edge modes, see Ref.~\cite{Fidkowski2010entanglement}. 

The correlation matrix spectrum for non-Hermitian systems was first studied in Ref.~\cite{herviou_entanglement_2019}. 
Since left and right eigenstates are not equivalent any more, there are in principle three distinct versions of Eq.~\eqref{eq:CA_gen}, using either the left-left, right-right, or left-right ground states. 
As argued in \cite{herviou_entanglement_2019}, the left-right choice is usually the one that behaves best and which we will use throughout the manuscript. 
Numerical simulations suggest that the topological features of the EOS for line gap phases are similar to those of hermitian gapped phases. 
In particular, one finds a one-to-one correspondence between the topological zero-energy states in the open system and the $\xi=1/2$ modes in the EOS. 
For a sublattice symmetric model with winding number $\nu$ there will be $2\nu$ topological zero energy modes --- $\nu$ on each edge --- and the same number of $1/2$ modes in the EOS.

Focusing on  two-band models with sublattice symmetry as in \eqref{eq:Hamiltonian}, an intuitive way of arguing for this is the following:
The correlation matrix in momentum space can be written as 
\begin{align}\label{eq:C_Q}
    C(k) &= \frac{1}{2}(\mathbf{1} + Q(k))
\end{align}
where $Q(k)$ is given by 
\begin{align}\label{eq:Q_gapped}
    Q(k)&=\mqty(0 & -\sqrt{f_1(k)/f_2(k)} \\ -\sqrt{f_2(k)/f_1(k)}&0) ,
\end{align}
using the explicit form of the left and right eigenvectors in \eqref{eq:eigenstates} and \eqref{eq:eigenstates_left}. 
We can choose the branch cuts of the square root in such a way that $Q(k)$ is continuous and differentiable. In addition, it is $2\pi$ periodic in $k$. 
Thus, we can interpret $Q$ as a  Hamiltonian with sublattice symmetry and compute the winding number relevant for line-gapped systems,
\begin{align}
    \nu&=\frac{1}{2\pi i}\int_0^{2\pi} dk \, \partial_k \ln\sqrt{f_1(k)/f_2(k)}\nonumber\\
    &= \frac{1}{2}\frac{1}{2\pi i }\int_0^{2\pi} dk \, \partial _k (\ln f_1(k) -\ln f_2(k))\nonumber\\
    &= \frac{1}{2}(\nu_1 +\nu_2).
\end{align}
Since the system is line-gapped, $\nu_1=\nu_2 =\nu$ and $Q(k)$ has the same winding number as the original Hamiltonian. 
Computing $C^A$ amounts to computing the real space Fourier transform of $Q$ in region A. The latter harbors $2\nu$ topological zero modes, resulting in $2\nu$ 1/2 modes in the EOS using Eq.~\eqref{eq:C_Q}. 

The situation for Hermitian critical and non-Hermitian pointgap phases is more complicated. 
An imminent problem is that it is not clear which ground state to use in \eqref{eq:CA_gen}. 
Numerical simulations show that for point-gapped systems, the EOS seems to harbor $\min(N_L,N_R)$ 1/2 modes, where $N_L$ and $N_R$ are the topological zero modes of a semi-infinite system \cite{ortega-taberner2022polarization}. 
In addition, these topological features were seen to be insensitive to the particular choice of ground state, while non-topological modes may be affected.
In our numerical simulations, the same seems to hold for critical systems, where now $N_L=N_R$ are the number of topologically protected edge modes on the left/right edge in the open boundary system. 
We will comment more on this in Section \ref{sec:EOS}, where we will also prove the equivalence of the EOS between Hermitian critical and non-Hermitian point-gapped phases for simple, yet non-trivial cases.

\section{Generalizing eigenenergies}\label{sec:generalization_1}

As we mention in the introduction, the aim of this paper is to show the connection in the topological features of critical Hermitian models and point-gapped non-Hermitian ones. 
In order to do that we first start with Hermitian models at a critical point and deform them to become non-Hermitian, showing that the topological features remain unchanged by this deformation. 
We will consider two distinct ways of deforming the Hermitian critical system. 
The first and simpler approach, discussed in this section, amounts to making eigenenergies  complex, while keeping the eigenstates unchanged.
Since we do not modify the eigenstates, the left and right non-Hermitian eigenstates are still equal to each other and one can use all the usual techniques in Hermitian quantum mechanics. 
To further simplify the discussion, we will only consider  critical systems where the gapless points are zeroes of order 1. 
Generalizing to higher orders is straightforward.
A less restrictive generalization, which also modifies the eigenstates,  is discussed in Section \ref{section:complex_momenta}.

For a given Hermitian  Hamiltonian, written in the eigenbasis, 
\begin{align}
 H=&  \sum_{k\mu} \varepsilon_{k\mu}\ket{\psi_{k\mu}}\bra{\psi_{k\mu}},
\end{align}
the eigenenergies are modified by adding the following perturbation:
\begin{align}\label{eq:gen_energies}
    H(g)=&  \sum_{k\mu} \varepsilon_{k\mu}(g)\ket{\psi_{k\mu}}\bra{\psi_{k\mu}},\nonumber\\
    =&
     \sum_{k\mu} (\varepsilon_{k\mu}+ig\partial_k \varepsilon_{k\mu}) \ket{\psi_{k\mu}}\bra{\psi_{k\mu}},
\end{align}
where $\varepsilon_{k\mu}$ are the energies of the Hermitian system, $\mu$ is the band index, and $\partial_k \varepsilon_{k,\mu} $ is by assumption non-zero. By construction, the eigenstates remain intact. 
As will be shown below, this generalization results in a point-gapped Hamiltonian. 
However, we first want to illustrate this generalization by the simplest possible example, assuming eigenenergies $\varepsilon_{k,\mu} = \pm \sin(k/2)$ with a single gapless point at $k=0$.
In order to construct the resulting complex bandstructure, it is easier to regard the model as an effective one-band model in an extended Brillouin zone $k\in [-2\pi,2\pi)$. 
The resulting complex energies are
 \begin{equation}
 \tilde{\varepsilon}_{k}(g) = \sin(k/2) + i\frac{g}{2}\cos(k/2),
 \end{equation}
 which is nothing else than the parametric equation of an ellipse in the complex plane. The system has therefore a point-gap around zero.
 More generally, zeroes of the hermitian model correspond to crossings of the imaginary axis (with finite imaginary part), while maxima and minima correspond to crossings of the real axis.

In terms of the Hamiltonian \eqref{eq:Hamiltonian}, the generalization in \eqref{eq:gen_energies} can be written as
\begin{align}\label{eq:Hamiltonian_energy}
H(k,g) =& \epsilon(k,g) \hat{h}(k) \cdot \boldsymbol{\sigma} \nonumber\\
=& [\epsilon(k) + ig\partial_k \epsilon(k)]\hat{h}(k) \cdot \boldsymbol{\sigma} 
\end{align}
with both $\epsilon(k)$ and $\hat h(k)$ continuous and differentiable. 
This can always be done for finite-range hopping Hamiltonians.  
Even though $\hat h(k)$ is not uniquely defined at a gapless point $k_c$, there is a consistent limit  
\begin{align}
	\lim_{\delta \rightarrow 0} \hat h(k_c+\delta) &= \lim_{\delta \rightarrow 0} \hat h(k_c-\delta) 
\end{align}
as long as $\epsilon(k)$ changes sign at $k_c$ (for a single zero).  
For explicit examples on how to choose $\epsilon(k)$ and $\hat h (k)$, we refer the reader to Appendix \ref{sec:examples}.

We now proceed to compute the winding number $\nu$ for the original, critical Hamiltonian, $H(k,0)$, using the conventions of \eqref{eq:Hamiltonian_energy}. 
A short calculation shows that $\nu$ only depends on $\hat h$, but not on $\epsilon$: 
\begin{align}\label{eq:winding_h}
  \nu_\epsilon&=\frac{1}{2\pi}\int_{()}dk\, \frac{(\partial_k h_x(k) ) h_y(k)-h_x(k) \partial_k h_y(k)}{ h_x(k)^2+h_y(k)^2} \nonumber\\
   & = \frac{1}{2\pi}\int_{()}dk\, \frac{\epsilon(k)^2\left(\hat h_x'\hat h_y-\hat h_x \hat h_y'\right)}{\epsilon(k)^2}\nonumber \\
   &\rightarrow\frac{1}{2\pi}\int dk\, \left(\hat h_x'\hat h_y-\hat h_x \hat h_y'\right) \,\mbox{ for } \epsilon\rightarrow 0
\end{align}
where we abbreviated $\partial_k \hat h_\alpha = \hat h_\alpha '$.
The last equality is valid as long as the integrand is regular, i.e. neither $\hat h$ or $\hat h'$ has singularities at the gapless points of the critical system. 
This is trivially satisfied for the finite-range hopping models usually considered.  
The resulting winding number is half-integer quantized, see \cite{verresen2020}. 
Repeating the same computation for the non-hermitian Hamiltonian, we find that the winding number 
\begin{align}\label{eq:winding_nh}
 \nu(g)  &= \frac{1}{2\pi}\int dk\, \frac{(\partial_k h_x(k,g) ) h_y(k,g)-h_x(k,g) \partial_k h_y(k,g)}{ h_x(k,g)^2+h_y(k,g)^2} \nonumber\\
 & = \frac{1}{2\pi}\int dk\, \frac{\epsilon(k,g)^2(\hat h_x'\hat h_y-\hat h_x \hat h_y')}{\epsilon(k,g)^2},\nonumber\\
 & = \frac{1}{2\pi}\int dk\, \left(\hat h_x'\hat h_y-\hat h_x \hat h_y'\right) 
\end{align}
is again independent of $\epsilon$ (and thus $g$) and equal to the last line of \eqref{eq:winding_h}. 
This shows that the generalization of the critical system to the non-hermitian one does not alter the winding number $\nu$. 

We can now continue taking a look at the second non-Hermitian winding number
\begin{align}
\label{eq:nu_prime}
    &\nu' = \frac{1}{2\pi}\oint_0^{2\pi}dk\, \partial_k\Im \log \det[h(k,g)],
\end{align}
which is equivalent to the phase winding of one of the energy bands. 
In order to use the form of Eq.~\eqref{eq:Hamiltonian_energy}, it is advantages to compute $2\nu'$ by extending the integration from $0$ to $4\pi$, thus allowing us to consider $\epsilon(k,g)$ as one of the energy bands.
In order to compute the winding, we now count the crossings at the positive imaginary axis (with an additional sign depending on the orientation). 
We will now show that the contribution of all gapless points to the winding number sums up, with only an overall sign depending on the sign of $g$. 
Let us fix the sign of $g$ to be positive and focus on one of the gapless points, denoted by $k_c$. 
If $\partial_k\epsilon(k)>0$, the crossing happens on the positive imaginary part from $\Re(\epsilon(k_c-\delta))<0$  to $\Re(\epsilon(k_c+\delta))>0$. 
If instead $\partial_k\epsilon(k)<0$, the crossing happens on the negative imaginary part from $\Re(\epsilon(k_c-\delta))>0$  to $\Re(\epsilon(k_c+\delta))<0$, i.e. it has the same orientation as above. 
Thus, the contributions of all gapless points simply add up with a positive sign. 
Switching the sign of $g$, switches the orientation. 
Thus, the winding number $\nu'$ is given by 
\begin{equation}\label{eq:nu'_result}
	\nu'= \rm{sign}(g)N/2,
\end{equation}
where the $\frac 1 2 $ is because our previous argument was made for $2\nu'$. 
Clearly, $\nu'$ itself is not well-defined in the $g\rightarrow 0$ limit, though its absolute value is and simply counts the number of gapless points.

\section{Generalization to complex momentum}
\label{section:complex_momenta}

So far we have shown a non-Hermitian generalization of critical systems that can be used to relate the topology of both models. However, the generalization considered,  modifying only the eigenvalues and not the eigenstates, is overly simplistic and artificial.
A more realistic generalization one can make is by performing an analytical continuation of the momenta in the Bloch Hamiltonian into the complex plane, $H(k)\rightarrow H(k,g)= H(k-ig)$. For a Hermitian model, a general Bloch Hamiltonian with sublattice symmetry can be written as
\begin{align}
\label{eq:bloch_hermitian}
    &H(k) = \mqty(0 & f \\ f^* & 0),
    &f= \sum_{n}t_n e^{ikn},
\end{align}
where $t_n$ are all possible hoppings. The non-Hermitian generalization reads
\begin{equation}
    H(k,g) = \mqty(0 & \sum_{n}t_n  e^{gn} e^{ikn} \\ \sum_{n}t_n^* e^{-gn} e^{-ikn} & 0).
\end{equation}
It  is equivalent to introducing non-reciprocity in the inter-cell hoping terms, which is one of the most common ways of introducing non-Hermiticity.

Before continuing, let us consider $g$ as a small perturbation,
\begin{align}
H(k,g) =& \boldsymbol{h}(k,g) \cdot \boldsymbol{\sigma} \nonumber \\
=& h(k,g)  \hat{h}(k,g) \cdot \boldsymbol{\sigma}\nonumber \\
\approx & \left[ h(k)  \hat{h}(k) + g (\partial_g h(k,g)|_{g=0})\hat{h}(k) \right. \nonumber \\
&\left. +  g h(k) \partial_g \hat{h}(k,g)|_{g=0}\right]\cdot \boldsymbol{\sigma}.
\end{align} 
Using that $\partial_g \boldsymbol{h}(k,g) = -i\partial_k\boldsymbol{h}(k,g) $,
\begin{align}\label{eq:complex_momenta_energies}
H(k,g) \approx & \left[ h(k) -i g \partial_k h(k) \right] \hat{h}(k)\cdot \boldsymbol{\sigma} \nonumber\\
& -i  g h(k) \partial_k \hat{h}(k)\cdot \boldsymbol{\sigma},
\end{align} 
we note that the first term is exactly the generalization considered in the previous section (up to the unimportant sign of $g$), but we now obtain a second term that also modifies the eigenvectors. Therefore, in the perturbative limit the two generalizations can be similar, but they are not identical. 

In contrast to the first generalization to non-Hermitian systems considered in Section \ref{sec:generalization_1}, it is now much easier to compute the winding numbers $\nu_{1/2}$ and then use their relation to $\nu$ and $\nu'$ \eqref{eq:winding2} to connect the result to the critical system. 
We begin by rewriting the expression for winding number $\nu_1$ in \eqref{eq:nu1/2} in the complex plane, employing  $z=e^{ik}$ as the holomorphic variable
\begin{align}\label{eq:nu_1_first}
	\nu_1&=\frac{1}{2\pi i}\oint_{|z|=1} dz \, \partial_z \ln[f(z e^{-g})]
\end{align}
and using $f(z)=\sum_n t_n z^n$, i.e. the appropriate expression for the critical system. 

We can absorb the scale factor by a change of variables, $z'=z e^{-g}$, and obtain
\begin{align}\label{eq:nu_1_second}
	\nu_1(g) = \frac{1}{2\pi i}\int_{\abs{z}'=e^{-g}} dz'\, \partial_z' \, \ln[f(z')] . 
\end{align}
which can now be obtained as $\nu_1 = Z_1-P_1$, where $Z_{1}(P_{1})$ are the zeros (poles) of $f(z)$  inside the circle $|z|=e^{-g}$ (times their multiplicity).  Note that, for the Hamiltonian considered here,
\begin{align}
    f(z) =& \sum_{n=-l}^r t_n z^n \nonumber\\
    =& \frac{P_{l+r}(z)}{z^l}
\end{align}
is a polynomial with $P=l$ poles at zero and $l+r$ zeros. For most of the commonly used model Hamiltonians, $l=0$ (e.g. the extension of the SSH model used in Appendix \ref{sec:examples} or the Hatano-Nelson model), but this is by no means a prerequisite. 

For $\nu_2$ we perform exactly the same steps, after substituting $k'=-k$ so that $f_2$ becomes a holomorphic, not anti-holomorphic function after expressing $z=e^{ik'}$: 
\begin{align}\label{eq:nu_2_first}
	\nu_2&=
	\frac {1}{2\pi i} \int_0^{2\pi} dk' \, \partial_{k'} \ln f(ze^g) \nonumber\\
	&=\frac{1}{2\pi i}\oint_{|z|=1} dz \, \partial_z \ln[f(z e^{g})]. 
\end{align}
We can again absorb the scale factor by a change of variables, but now we set $z'=ze^{g}$ instead.
Consequently, $\nu_2$ is given by $\nu_2 = Z_2-P_2$, where $Z_{2}(P_{2})$ are now the zeros (poles) of $f(z)$  inside the circle $|z|=e^{g}$.
 
 For infinitessimal $g>0$, $\nu_1$ counts (Z-P) for all zeroes/poles that lie strictly within the unit circle, $\nu_1 = \tilde{\nu}$, while $\nu_2$ even counts the zeroes that lie on the unit circle.  
Assuming we don't go through any phase transition when enlarging $g$, the difference of the two winding numbers is simply the number of zeros of $f(z)$ on the unit circle times their multiplicity,
\begin{align}\label{eq:nu1_zeroes}
	\nu_1(g) - \nu_2(g) =& N\nonumber  \\
	=& 2\nu'.
\end{align}
When the underlying critical system has only single zeroes, this is consistent with our previous results \eqref{eq:nu'_result}. \footnote{In case the critical system has zeroes with higher multiplicity, the two generalizations to non-hermitian systems are \emph{not} equivalent, as \eqref{eq:nu'_result} only depends on the number of zeroes, but not their multiplicity. This is a direct consequence of the generalization in Section \ref{sec:generalization_1} being too simplistic.  }
Interestingly, as long as we look at phase transitions where zeros do not move simultaneously into and out of the unit circle, $\nu_1,\nu_2$ also correspond to the winding numbers of the gapped Hermitian phases surrounding the gapless point. 
However, this interpretation of $\nu_1,\nu_2$ depends on our particular generalization from critical to non-Hermitian systems and will not hold in general. 

In order to illustrate this, we  consider a very small anti-Hermitian perturbation to our Hamiltonian, such that
\begin{align}
	&f_1(k)= f(k) + \delta \nonumber \\
	& f_2(k) = f(k) - \delta^* .
\end{align}
Nonzero $\delta$ will generically move the complex zeroes either into or out of the unit circle. Let us  first  consider $\nu_1$, and observe that since $f(z)$ has a zero on the unit circle, we can write $f_1(z)$ as 
\begin{align}
	f_1(z) = A(z)(z-z_0)+\delta,
\end{align}
where $z_0$ denotes the position of the gapless point. 
Given that $\delta$ is small, we can approximate $A(z)$ by a constant in the environment of $z_0$, yielding 
\begin{align}
	f_1(z) \approx A(z_0) \left( z-z_0 + \delta/A(z_0) \right)
\end{align}
for $z\approx z_0$. 
Similarly, for the other winding number we have
\begin{align}
	f_2(z) \approx A(z_0) \left( z-z_0 - \delta^*/A(z_0) \right)
\end{align}
From these expressions we can now deduce how the zeroes move. 
In the case where $\delta$ is real, the zero of one of the functions moves inside the unit circle while the other one moves out, depending on the sign of $\delta *(z_0^*A(z_0)^*+z_0A(z_0))$. This results in $\nu_1,\nu_2$ being the winding numbers of the gapped neighbouring phases of the gapless point. That is no longer the case for purely imaginary $\delta$, as both zeros move in the same direction.

\section{Entanglement Occupancy Spectrum}\label{sec:EOS}
In this section, we show some rigorous results on the EOS for the non-Hermitian generalizations considered above. 

\paragraph{EOS for generalizing eigenenergies}
In this simple case it is trivial to see that the EOS is independent of $g$, the non-Hermitian parameter. This is because the EOS only depends on the occupied eigenstates, which are not modified by this particular generalization. If we choose to occupy the same eigenstates, independently of $g$, then the Hermitian and non-Hermitian models will share the same EOS. Furthermore, since the real part of the energies is also independent of $g$, the common choice of occupying the states with $\Re[\varepsilon_{k\mu}]<0$ will lead to the same EOS.

When extending the critical model to non-Hermitian phases, we note that $\tilde \nu$ in Eq.~\ref{eq:number_top_edge_modes} becomes identical to $\nu_1$ ($\nu_2$) for sufficiently small $g>0$ ($g<0$). The second important quanity, the number of gapless points on the unit circle $N$, is encoded in $|\nu'|=|\nu_1-\nu_2|/2$. 
For the non-Hermitian phases that are of current interest, $\nu_{1/2} \geq 0$, which explains their identification with the number of left/right topological edge states in Ref.~\cite{Yin2018}. 
However, this identification will fail if one or both of the winding numbers are negative, and one needs to revert to similar arguments as used in Ref.~\cite{verresen2018} to derive the correct number of topological edge modes. 
In general, extensive numerical simulations suggest that the number of topological $1/2$ states in the EOS is equal to  $\min(N_L, N_R)$, where $N_{L/R}$ are the number of edge modes in the semi-infinite chain, see Ref.~\cite{ortega-taberner2022polarization}. 
In the limit of the Hermitian critical system, this reduces to Eq.~\eqref{eq:number_top_edge_modes}. 
Note that even though the second non-Hermitian winding number $\nu'$ does not have a well-defined Hermitian limit, its absolute value, $|\nu'|=N/2$, still carries topological information.

\paragraph{EOS for generalizing to complex momenta}

Making the momenta complex, $k\rightarrow k-ig$, modifies the eigenenergies in a similar way as for our first generalization (see e.g. \eqref{eq:complex_momenta_energies}), but it will also modify the eigenstates. 
So it is far from obvious that the EOS will remain qualitatively unchanged by this generalization. 

Instead of the correlation matrix P, we will consider the matrix 
\begin{align}\label{eq:Qgen}
    Q = 2P-I,
\end{align}
see Eq.~\eqref{eq:Q_gapped}.
The spectrum of the subsystem matrix $Q^A$ is equivalent to the EOS, with the virtual topological states having zero eigenvalue. 

For critical systems, the ground state is degenerate and $Q$ generally depends on the particular ground state chosen. 
For systems with only one gapless point at $k_c$, there is a well defined procedure to choose the ground state by filling either the $+$ or the $-$ band from \eqref{eq:eigenstates} for  $k\in (k_c, k_c + 2\pi]$. \footnote{This choice ensures that the Resta polarization behaves consistent with the $T\rightarrow 0 $ limit of finite-temperature computations \cite{Balabanov_2022}. }
Using this choice of ground state, the momentum space representation of $Q$ becomes  
\begin{align}\label{eq:Qk}
    Q(k) = \mqty(0 &- e^{i \phi(k)} \\ -e^{-i \phi(k)}),
\end{align}
where $e^{i \phi(k)} = \sqrt{f(k)/f^*(k)}$.
Note that $\phi(k)$ is \emph{not} $2\pi$-periodic, in contrast to Hermitian gapped or non-hermitian line-gapped phases, but $4\pi$-periodic, since Eq.~\eqref{eq:eigenstates} implies that $v_+(k+2\pi) = v_-(k)$.  
After the non-Hermitian generalization, the $Q$-matrix becomes 
\begin{align}
    Q(k,g) = \mqty(0 & -e^{i \phi(k-i g)} \\ -e^{-i \phi(k-ig)}).
\end{align}

We now assume that $\phi$ has a special form and can be written as $\exp[i\phi(k)] = \exp(ik \alpha)$. This is, for instance, possible for the critical SSH chain with $\alpha=1/2$. 
In this case,  the full $Q^A$ matrix is given by
\begin{align}
    Q^A_{xy}(g) =& \oint \frac{dk}{2\pi} e^{ik(x-y)}\mqty(0 & e^{i \phi(k-i g)} \\ e^{-i \phi(k-ig)}) \nonumber\\
    =& \oint \frac{dk}{2\pi} e^{ik(x-y)}\mqty(0 & e^{i \alpha k} e^{g\alpha} \\  e^{-i \alpha k} e^{-g\alpha}) \nonumber\\
    =&\oint \frac{dk}{2\pi} e^{ik(x-y)} 
   \mqty(e^{g\alpha/2} & 0 \\ 0 & e^{-g\alpha/2})\nonumber\\ &\times 
    \mqty(0 & e^{i \alpha k} \\  e^{-i \alpha k} ) \mqty(e^{-g\alpha/2} & 0 \\ 0 & e^{g\alpha}/2)\nonumber\\
    =& U(g)Q^A_{xy}(0) U(g)^{-1},    
\end{align}
where $x,y$ are restricted to lie in A.  
This is a similarity transformation that leaves the whole spectrum invariant, not just the topological states that one is interested in. 

\begin{figure}
    \centering
    \includegraphics[width=\columnwidth]{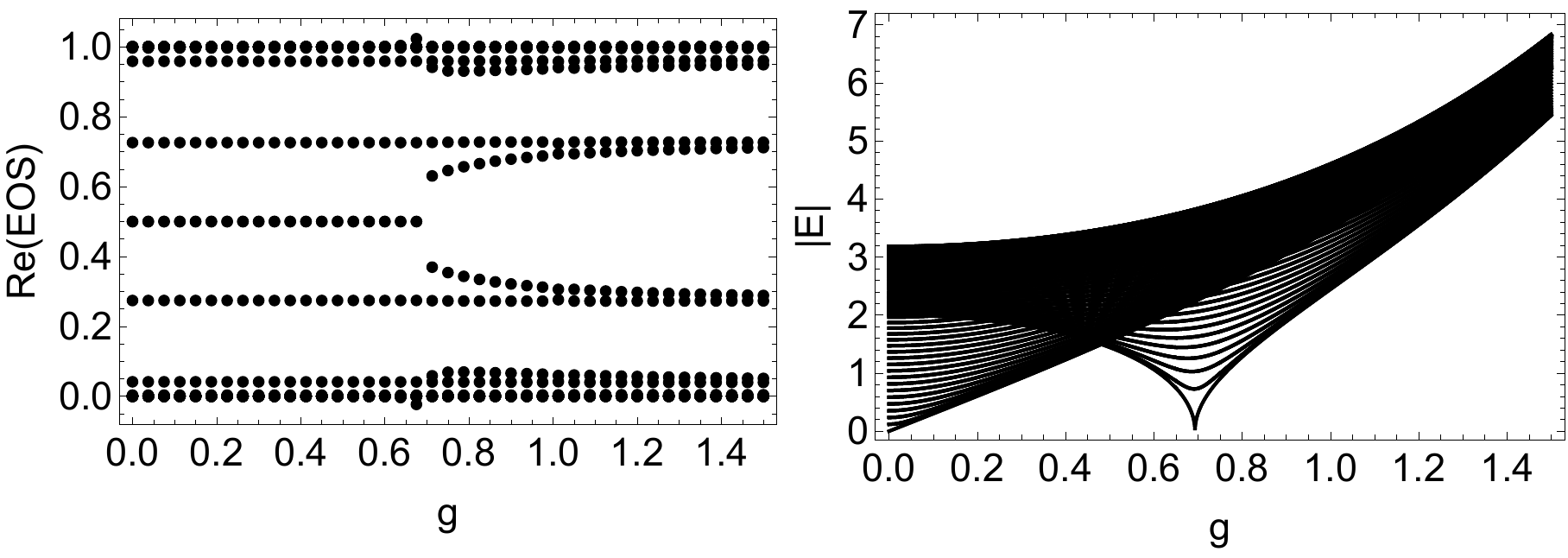}
    \caption{left: The real part of the EOS as a function of $g$, when generalizing the critical Hamiltonian defined by $f(k)= 1+e^{ik}-2e^{2ik}$, see Eq.~\eqref{eq:Hamiltonian}. The EOS remains qualitiatively unchanged until $g=\ln 2$, where a phase transition to a trivial phase occurs, see right figure.  }
    \label{fig:EOS}
\end{figure}

In general, $\phi$ is not of the special form assumed above and the two spectra will not be equal to each other. Extensive numerical simulations suggest that the 1/2 modes do remain robust, however. Moreover, in models with a finite range, i.e. the ones most commonly used as model systems, we observe only small deviations in the spectra, i.e. they remain \emph{qualitatively} the same.
We exemplify in Fig.\ref{fig:EOS}, using the critical system  defined by $f(k)= 1+e^{ik}-2e^{2ik}$, see \eqref{eq:Hamiltonian}. 

Proving the robustness, however, turns out to be difficult. 
One avenue might be to use the same techniques as in Ref.~\cite{Monkman2022} to construct exact eigenstates of $C^A$, and link their number to the relevant winding number in the thermodynamic limit. 
Constructing the form of the $C^A$ eigenstates for both non-Hermitian point-gapped systems and critical systems with only a single gapless point is straightforward.
However, this only tells us how eigenstates look if they exist, but not how many of those exist. 
For the latter, we  need to generalize the arguments by \cite{Monkman2022} to relate their number to the winding number in the thermodynamic limit -- a highly non-trivial task. 

Another avenue is to construct zero energy eigenstates of $Q^A$ for finite $g$, using their expression from $g=0$. 
This would not prove the correspondence between the number of topological zero-energy edge modes and $1/2$ modes of $C^A$, but only establish that the topological modes of $C^A$ survive the non-Hermitian generalization. 
This seems to be more tractable, but is still work in progress. 

\section{Conclusion}\label{sec:conclusion}

In this manuscript, we have proven analytically that the topological invariants characterizing Hermitian critical systems are the same as those of two distinct generalizations to non-Hermitian point-gap systems.
While one of the generalizations also leaves the EOS --- a useful tool for characterizing topology --- invariant, the other one in general does not. 
Despite that, numerical simulations strongly suggest that the topological features remain unchanged. 
We have proven this explicitly for a simple, yet non-trivial model. 
The close relation between Hermitian critical and non-Hermitian point-gapped phases also lends some physical interpretation to some properties of the latter, in particular explaining the number of topological modes in the EOS. 

Our results  might be utilized in at least two different ways: we can regard the non-Hermitian generalizations discussed in Sections \ref{sec:generalization_1} and \ref{section:complex_momenta} as a regularization of the critical phase, which might prove useful in studying critical systems. 
It also allows us to compute quantities for non-Hermitian models using Hermitian physics
and to address questions regarding critical Hermitian systems using the well-studied topological classification for non-Hermitian systems.

During the completion of this manuscript, another work was put forward \cite{hsieh2022relating} that deals with the relation of non-Hermitian to Hermitian critical systems.
However, in \cite{hsieh2022relating} the authors relate Hermitian critical systems to non-Hermitian critical systems.
An important aspect of their construction is that the non-Hermitian energy spectrum is real (and gapless). 
Thus, their construction is distinct from ours, even though many of the features are similar. 
In particular, also for point-gap phases one obtains an entanglement entropy scaling linear in sub-system size, implying critical behavior. 

\emph{Acknowledgements---}
We want to thank Lukas R\o dland and Eddy Ardonne for useful discussions. 
The research in this grant was
supported by the Swedish Research Council under grant
no. 2017-05162 and the Knut and Alice Wallenberg foundation under grant no. 2017.0157.

\begin{appendix}
\section{Explicit examples}\label{sec:examples}
\begin{figure}[t]
	\centering
	\includegraphics[width=\columnwidth]{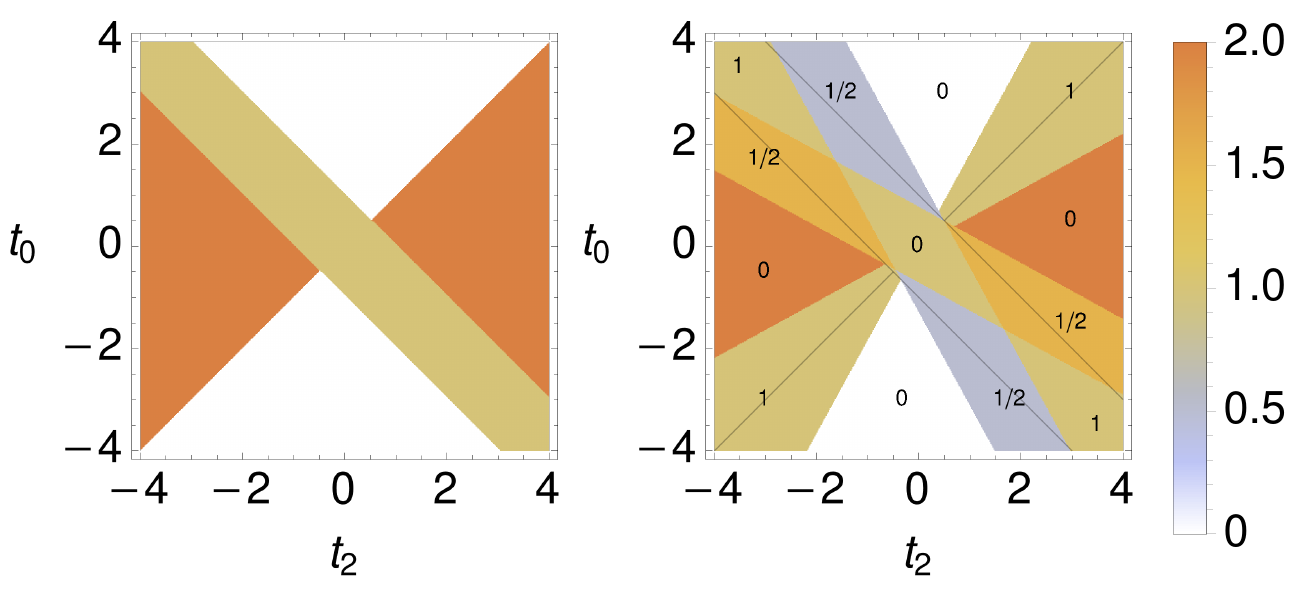}
	\caption{Phase diagram for Hamiltonian \eqref{eq:model_hamiltonian} with $t_1=1$ for  $g = 0$ (left) and  $g=0.2$ (right). The color code indicates $\nu$, while the numbers indicate $\nu'$ (only non-zero for $g\neq 0$). }
	\label{fig:phase_diagram}
\end{figure}
In this section, we want to exemplify some of the statements of the main text. Throughout, we use a simple model Hamiltonian given by
\begin{align}\label{eq:model_hamiltonian}
    H &= {\boldsymbol h(k)}\cdot \sigma \qquad \mbox{ with } \nonumber\\
    h_x (k)&= \gamma_0 +\gamma_1 \cos(k)+\gamma_2 \cos(2k)\nonumber\\
    h_y (k) &= \gamma_1 \sin(k)+\gamma_2 \sin(2k)
\end{align}
which implies that 
\begin{align}
    f (k) &= \gamma_0 +\gamma_1 e^{ik} +\gamma_2 e^{2ik}. 
\end{align}
The phase diagram of the hermitian system is shown in \ref{fig:phase_diagram}(a). 
We will focus our analysis on four critical points given by 
\begin{align}
  \{\gamma_0,\gamma_1,\gamma_2\} 
 & =& & (i) &&\{0,1,-1\}  &(\nu:\, 1\rightarrow 2),\nonumber\\
  &=& &(ii)& &\{-1, 1, 0\} &(\nu:\, 0\rightarrow 1),\nonumber\\
  & =& & (iii)&& \{-1, 1, -1\} &(\nu:\, 0\rightarrow 2),\nonumber\\ 
   & =& & (iv)&& \{-\frac 1 2, 1, -\frac 1 2\} &(\mbox{tricritical}).
\end{align}
Without loss of generality we assume $g>0$ in the following.

\paragraph{Generalizing eigenenergies}
Let us first consider the first method, described in Section \ref{sec:generalization_1}, on how to connect these critical points to non-Hermitian point-gap phases. Below, we give the analytical expressions for $\epsilon(k,g)$ as well as $\hat h$. 

(i) $\nu: 1\rightarrow 2$: 
\begin{align}
    \epsilon(k,g)&= 2 \sin(k/2) + i g \cos(k/2)\nonumber\\
    \hat h_x (k) &=\sin[(3 k)/2]\nonumber\\
    \hat h_y(k)&= \cos[(3 k)/2]
\end{align}

(ii) $\nu: 0\rightarrow 1$: 
\begin{align}
    \epsilon(k,g)&= 2 \sin(k/2) + i g \cos(k/2)\nonumber\\
    \hat h_x (k) &=-\sin(k/2),\nonumber\\
    \hat h_y(k)&= -\cos(k/2)
\end{align}
(iii) $\nu: 0\rightarrow 2$: 
\begin{align}
    \epsilon(k,g)&= 1-2 \cos (k)+2 i g  \sin (k)\nonumber\\
    \hat h_x (k) &=\cos(k),\nonumber\\
    \hat h_y(k)&= -\sin(k)
\end{align}
(iv) tricritical: 
\begin{align}
    \epsilon(k,g)&= 1- \cos (k)+ i g  \sin (k)\nonumber\\
    \hat h_x (k) &=\cos(k) \nonumber\\
    \hat h_y(k)&= -\sin(k)
\end{align}
Note first that whenever there are an odd number of gapless points, $\epsilon(k)$ cannot be chosen as   $2\pi$ periodic, but must be $4\pi$ periodic. This is intimately related to $\hat h (k+2\pi)=-\hat h(k)$ and, thus, a half-integer quantized winding number. The winding numbers of (i) and (ii) are $3/2$ and $1/2$ respectively. 
For case (iii) and (iv), we note that both $\epsilon$ and $\hat h$ are $2\pi$-periodic. The corresponding winding number is 1. 
On first glance, it may be suprising that the winding number of the tricritical point is at all well-defined. 
It is, however, a fact that appears again in our second non-Hermitian  generalization.
In the latter, the tricritical point connects to the same point-gapped phase as (iii). 

\paragraph{Generalizing to complex momenta}
We now consider the non-Hermitian generalization of critical systems, obtained by letting the momenta become complex as $e^{ik}\rightarrow e^{ik-g}$. 
The resulting phase diagram is shown in Fig.~\ref{fig:phase_diagram}, with the color code indicating $\nu$ and the numbers indicating $\nu'$. 
Note that increasing $g$ turns the critical line into point-gap phases, while gapped phases smoothly evolve to line-gapped phases (with $\nu'=0$) as expected \cite{PhysRevX.9.041015}.

For (i) $\nu: 1\rightarrow 2$ and using $z=e^{ik}$, $f_1$ is given by
\begin{align}
    f_1(z)&=e^{-g} z - e^{-2 g} z^2 = e^{-g} z(1-e^{-g}z),
\end{align}
which has single zeroes at $z=0$ and $z=e^{ig}$. 
Thus, the second zero has moved outside the unit circle and the resulting winding number in Eq.~\eqref{eq:nu_1_first} evaluates to 1. 
For $f_2$ we instead find
\begin{align}
    f_2(z)&=e^{g} z - e^{2 g} z^2 = e^{g} z(1-e^{g}z)
\end{align}
with $z=e^{ik'}=e^{-ik}$, which has single zeros at $z=0$ and $z=e^{-ig}$. 
Thus the  zero has moved inside the unit circle and the resulting winding number, see Eq.\eqref{eq:nu_2_first}, is 2. 

For case (ii), $f_1/f_2$ instead become 
\begin{align}
    f_{1/2}&= -1 +z e^{\mp g}, 
\end{align}
which when compared to (i) above lacks the single zero at $z=0$, thus yielding $\nu_1=0$ and $\nu_2=1$ when evaluating the winding number integrals. 

For case (iii), 
\begin{align}
    f_{1/2}&= -1 + z e^{\mp g} - z^2 e^{\mp 2g}\nonumber\\
    &=\left(z- e^{-\pi i/3} e^{\mp g}  \right) \left(z-e^{i\pi/3} e^{\mp g}\right),
\end{align}
thus both zeroes move simultaneously either inside or outside the unit circle yielding $\nu_1=0$ and $\nu_2=2$. 

Last but not least, for case (iv) one finds
\begin{align}
    f_{1/2}&= -\frac 1 2 + z e^{\mp g} - \frac 1 2 z^2 e^{\mp 2g}\nonumber\\
    &=\left(z-  e^{\mp g}  \right)^2,
\end{align}
which is completely equivalent to case (iii), except that both zeroes now sit at the same point. Again, the winding numbers are given by  $\nu_1=0$ and $\nu_2=2$. 
In all these cases, the gapless point(s) for each winding number move either all inside or all outside, thus allowing us to identify $\nu_1$ and $\nu_2$ with the gapped phases on each side of the  critical point.

\end{appendix}
\bibliography{references}
\end{document}